\documentclass[pra,twocolumn,superscriptaddress,floatfix]{revtex4}
\usepackage{hyperref}
\usepackage{graphicx} 
\usepackage{amsmath}
\usepackage{float}
\usepackage[dvips]{epsfig}
\newcommand{\beq}{\begin{equation}}
\newcommand{\eeq}{\end{equation}} 
\newcommand{\beqa}{\begin{eqnarray}}
\newcommand{\eeqa}{\end{eqnarray}}
\newcommand{\ba}{\begin{array}}
\newcommand{\ea}{\end{array}}

\begin{document}

\title{Controlling quantum vortex dynamics and vortex-antivortex
annihilation in Bose-Einstein condensates with optical lattices}

\author{Francesco Ancilotto}
\affiliation{Dipartimento di Fisica e Astronomia ``Galileo Galilei''
and CNISM, Universit\`a di Padova, via Marzolo 8, 35122 Padova, Italy}
\affiliation{ CNR-IOM Democritos, via Bonomea, 265 - 34136 Trieste, Italy }

\author{Luciano Reatto}
\affiliation{ Dipartimento di Fisica, Universit\`a degli Studi di Milano, 
via Celoria 16, 20133 Milano, Italy  }

\begin{abstract} 

Superfluids with strong spatial modulation can be experimentally
produced in the area of cold atoms under the influence of optical
lattices. 
Here we address $^{87}$Rb bosons at T=0 K in a flat geometry
under the influence of a periodic potential with the Gross-Pitaevskii
theory. 
The statics and dynamics of vortex excitations are
studied in the case of one dimensional (1D) and of two dimensional
(2D) optical lattices, as function of the intensity of the
optical lattice. 
We compute how the vortex energy
depends on the position of its core and the
energy barrier that a vortex has to surmount in order to move in the
superfluid. 
The dynamics of a vortex dipole, a pair of vortices of
opposite chirality, differ profoundly from the case of 
a uniform superfluid. In the 1D case, when parallel
ridges of density are present, the dynamics depends on the positions
of the two vortices. If they are in the same channel between 
two ridges, then the two vortices
approach each other until they annihilate each other in
a short time. If the two vortices are in distinct channels the dipole
undergoes a rigid translation but with a
velocity depending on the intensity of the optical lattice and this
translation velocity can even change sign with respect to the case
of the uniform superfluid. Superimposed on this translation an
oscillatory motion is also present. A superposition of translation
along a channel and an oscillation is also found with a single vortex
when the system is bounded inside a circular trap. These oscillatory motions
can be both longitudinal, i.e. along the channel, as well as
transverse. In all cases the transverse motions are one-side, in
the sense that the vortex core never crosses the equilibrium
position nearest the starting position. In the case of the 2D
lattices we study (square, triangular and honeycomb), the two
vortices of a dipole move mainly by jumps between equilibrium
positions and approach each other until annihilation. This behavior
has some similarity with what has been found for a vortex dipole in
the supersolid state of dipolar bosons.
We show that a rapid rump-down of the
optical potential improves the visibility of the density holes at the vortex core.

\end{abstract} 
\date{\today}

\maketitle

\section{Introduction and summary}

One of the key properties of a superfluid is the presence of
quantized vorticity. In the case of bosons this has been
experimentally verified in superfluid $^{4}$He \cite{donnelly} and in Bose-Einstein
condensates (BEC) of cold atoms \cite{fetter}. A quantum vortex is an
excitation of the system with quantized circulation and
angular momentum and usually such excitation is studied in a
superfluid that is uniform or with an inhomogeneity that is weak on the
scale of the healing length of the superfluid. At present
there is the possibility of studying vortices in spatially
strongly inhomogeneous systems, both in terms of the spatial
scale as well as in terms of large contrast between the
maximum and the minimum densities. In the realm of cold atoms
it is possible to generate a periodic potential (optical
lattice, OL) acting on the atoms and this potential is obtained by
standing waves of suitable crossing light beams \cite{olattice}. The period of
the optical lattice can be comparable to the healing length
and the local density can vary even by orders of magnitude. A
recent experiment \cite{tao} has verified the theoretical prediction \cite{leg70}
that even at the lowest temperatures the superfluid
fraction of the superfluid is less than unity due to the
induced density modulation.

Another system that is
predicted to be a spatially strongly inhomogeneous superfluid
is a submonolayer of $^{4}$He adsorbed on a substrate of
fluorographene, a sheet of graphene decorated by fluorine
atoms. The adsorption potential of a He atom on this substrate
is strongly corrugated but not so much to cause localization
of the atoms like on graphite. Theory \cite{moroni} shows that the
system is superfluid and the local density has a very large
excursion in space. A supersolid (SS) offers another example
of a strongly inhomogeneous superfluid. In a SS the
inhomogeneity arises from a spontaneous broken symmetry of the
translational invariance. Such SS state has been found in cold
atom systems, for instance in dipolar bosons like $^{164}$Dy \cite{lchomaz}.
Again the local density can vary by a large amount and the
healing length is comparable with the lattice parameter of the
SS. 

Different ways have been
devised to create vortices in BEC superfluids, such as phase imprinting \cite{isoshima},
by dragging obstacles \cite{neely},
by means of artificial gauge fields \cite{jlin}, and more recently
using a versatile, deterministic  2D vortex collider in homogeneous atomic superfluids \cite{kwon}.

Complex 
dynamics of vortices under the action of the optical lattice in a trapped BEC sample
have been unveiled in a theoretical study of vortices within the two-dimensional GP equation with
the OL and magnetic trap \cite{kevrek}, where it was found that
depending on the phase of the OL relative to the
parabolic magnetic trap, it is possible either to trap the vortex at the center
of the trap, or the vortex moves along an unwinding spiral, towards the periphery of the trap.

The interplay of lattice physics and rotation physics 
was studied by calculating the vortex-lattice structures near
a Mott transition \cite{goldbaum}.
A single BEC loaded in a rotating OL can show
a rich variety of vortex structural transitions \cite{hpu}

There have been predictions of novel types of  
vortex states of repulsive BECs 
confined by a shallow optical lattice
(matter-wave gap vortices), which are spatially
localized and dynamically stable in 2D
as well as in 3D optical lattices \cite{kivshar}.

A somewhat related topics (vortex solitons in BEC stabilized by optical lattices)
have been adressed in Ref. \onlinecite{malomed}.

Interestingly, vortex dynamics in non-homogeneous systems can be
linked to "glitches" in rotating pulsars, i.e. a sudden speed up
of the spinning stars, occurring at random intervals.
These events are believed to be a manifestation 
of the presence of a superfluid component in the stellar interior, the glitch
occurring when many vortices jump from the inside of the
star to the solid crust, transferring angular momentum and thus speeding
it up. It has been shown recently, using cold atom experiments as 
analogs of neutron stars, that this requires
simultaneous
crystalline and superfluid phases, i.e. a supersolid state \cite{epoli}.

A subject that apparently has received very little attention 
is the behavior of vortex dipoles. Such configurations have been experimentally obtained
in BEC in the work of Refs. \onlinecite{neely}.
However topics such
as the interaction of such dipoles and the ensuing dynamics are still unexplored.
Our study provide an insight into these issues, and how they are modified
by optical lattice.

All this gives a
motivation for studying vortices in strongly inhomogeneous
superfluids, the aim of the present paper. We show how
strongly the dynamics of vortices and the lifetimes of pairs
of vortex-antivortex are affected by an optical potential. The
experimental realization of generation of vortex dipoles in a
suitable trap for cold atoms \cite{kwon} open the possibility of
experimental verification of our predictions.

Vortices in the SS state of matter in dipolar bosons have been 
addressed \cite{gallemi,casotti,gallemi1,Anc21} in recent years and some relevant phenomena
have been uncovered, like a reduced angular momentum
associated to the quantum of circulation \cite{gallemi}. A peculiar
behavior has been found \cite{Anc21} in the dynamic of a vortex dipole,
i.e. a pair of vortices of opposite chirality, in a SS. In a
uniform superfluid, when the distance $l$ between the two
vortices is larger than the healing length $\xi $, the pair is a
stable entity and moves with a constant velocity as a rigid
body in a direction perpendicular to the vector $\vec l$ joining the
two vortices. Contrary to this, in dipolar bosons in a SS
state it was found that the two vortices moves by jumps
between equilibrium sites in the lattice and approach each
other until the two vortices annihilate themselves in a short
time. At the basis of this behavior is a  
basic property of a vortex in a SS: it is
energetically favorable to have its core at a discrete set of
positions, the locus of minimum density. The natural question
is if such properties are specific of a SS state or if they are 
generic ones when a density modulation is present, whatever is
its origin. To answer this question we address in the present
paper the study of bosons in an external periodic potential
like that produced by light standing waves.
We study the ground state and vortices of the standard model
of BEC, point-like bosons with a contact interaction, in the
case of $^{87}$Rb with the Gross-Pitaevskii equation in an external
periodic potential $V_{ext}(\vec r)$. We consider the case of
modulation in one dimension (1D) and in two dimensions (2D) for
three lattices: square, triangular and honeycomb.

The system is subject to periodic boundary conditions
in all three space directions, and translational invariance 
along the z-direction perpendicular to the lattice potential plane is assumed.
The length in the z direction is of order of the
healing length of the superfluid so that the system can be
considered as a quasi-two dimensional system.

In the studied range of intensity of our optical potentials we find
that the bosons are in a superfluid state with a reduced
superfluid fraction. The local density is a periodic function
of position reflecting the symmetry of the optical potential
and the excursion between maxima and minima can be very large,
depending on the strength of the modulation of $V_{ext}(\vec r)$. In a
uniform superfluid the vortex excitation energy $\Delta E_v$ does
not depend on the position of its core. On the contrary, in the presence of the optical 
potential $\Delta E_v$ depends on the position of its core and it is a
strong function of the local density. The minima 
of $\Delta E_v$ are at the
positions of the minima of the local density, i.e. at the
maxima of $V_{ext}(\vec r)$, consistently with the observation
of pinning of the vortex at the low density site of an OL was observed \cite{duine}.
Therefore $\Delta E_v$ is a periodic function
of position and this has a dramatic effect on the dynamics of
vortices. The flow field of a vortex is strongly deformed
from the circular shape of the uniform case and we
characterize the vortex excitation energy and its structure as
function of the amplitude of the optical potential. 

The
dynamics of a vortex dipole is also strongly affected by the presence
of the optical potential.

For a 2D potential, where the
equilibrium positions of the vortex are isolated points, the
dipole moves mainly by jumps between equilibrium sites
approaching each other until the two vortices annihilate
themself in a short time. In our theory no thermal or stochastic 
effects are present and these jumps are manifestations 
of tunneling of the vortices between equilibrium sites.

The behavior is different in the
case of a 1D modulation where the equilibrium positions for a
vortex form a series of parallel lines in the x-y plane.

If the two vortices of
the dipole are located in the same channel they do not translate
but they move one against the other until they annihilate and
the excitation energy goes into phonons of the superfluid. If
the two vortices are located in adjacent channels they move along
these channels and the motion is a composition of
a uniform translation and of an oscillation. In all cases the
translation velocity of the pair differs from that expected
for a uniform superfluid and it can even invert the direction
of motion for large amplitude of the optical potential.
A combination of translation and oscillation is found also for a
single vortex in a trap in presence of a 1D OL.
The frequency of this oscillation is the same as in the case of a
vortex dipole moving along two neighboring channels. Therefore this oscillatory
motion seems to be an intrinsic and novel character of the motion of a vortex
moving along a channel. Notice that this oscillatory motion is quite different
from that of a massive particle around a minimum energy position because the
vortex oscillation is one-side, i.e. the vortex never crosses the position of the minimum
energy and it remains on the side from which it started.

The local density goes to zero at the
position of the vortex core. In the case of a strong
modulation experimentally it can be difficult to detect the
presence of a vortex because there is a small contrast between
the vanishing density of the vortex core and the small value
of the density at the minima of the modulated system. We show
that by starting from a state with a number of
vortices in the modulated superfluid, such vortices remain in
the homogeneous superfluid after that the optical potential is
suddenly removed, allowing for an easier visual detection
due to the increased contrast between the vanishing density at the core positions 
and the density of the surrounding, almost homogeneous phase.

The content of the paper is as follow. In Sect.II the theory
and the computational method are described. In Sect.III the
ground state, the vortex state and the dynamics of a vortex
pair are studied in the case of 1D optical potential.
In this Section we study also
the dynamics of a vortex when the system in 
confined in a circular trap in the x-y plane.
The case
of 2D optical potentials is studied similarly in Sect.IV. In
Sect.V we study the time evolution of a confined system after
the optical potential is removed, as a way to directly image vortex
positions in these highly inhomogeneous systems. 
Our conclusions are contained in Sect.VI.

\section{Method}

The Gross-Pitaevskii (GP) energy functional for the Bose system reads

\begin{widetext}
\begin{equation}
E = \int d{\bf r} \, \left[{\hbar^2 \over 2m}|\nabla \psi({\bf r})|^2  
+ V({\bf r}) \rho({\bf r})\right]\,
+\frac{1}{2}g\int d{\bf r} \, \rho ^2({\bf r}) \,
\label{eq:energyfun}
\end{equation}
\end{widetext}
where $V({\bf r})$ and $\rho({\bf r})=|\psi ({\bf r})|^2$ 
represent the external potential and the boson number density, respectively.
The coupling constant is $g=4\pi  a_s \hbar^2/m$, $m$ being the atomic mass.
The number density $\rho $ is normalized such
that $\int _V \rho({\bf r})\,{\rm d}{\bf r} =N$ where $N$ 
is the total number of atoms.
As model system we take $^{87}$Rb atoms and the scattering
length $a_s$ describing the (repulsive) Rb-Rb interaction
is $a_s = 100.4\,a_0$ \cite{marte} ($a_0$ being the Bohr's radius).

Minimization of the action associated to Eq.~(\ref{eq:energyfun}) leads to
the following Euler-Lagrange equation (GP equations) 

\begin{widetext}
\begin{equation}
i \hbar {\partial \psi (\vec r,t)\over \partial t} =
\left[-{\hbar^2 \over 2m}\nabla ^2 + V(\vec r) + g\rho (\vec r,t)\right]\psi (\vec r,t)\equiv H\psi (\vec r,t)
 \, ,
\label{eq:gpe}
\end{equation}
\end{widetext}

When steady states are studied the left hand side of Eq.\eqref{eq:gpe} is replaced by
$\mu \psi (\vec r,t)$ where $\mu $ is the chemical potential.
The numerical solutions of Eqs. \eqref{eq:gpe} provide 
the time-evolution of the $^{87}$Rb system with arbitrary $N$ 
in three-dimensions. 
The same equation in imaginary time
allows us to obtain stationary state solutions starting from a
suitable initial function. 
$\mu $ is determined so that the desired value of $N$ is
achieved.
We consider an external potential
depending only in $x$ and $y$. 

In our computations the length
$L_z$ of the computation box is much smaller of the other sides.
Under this condition we simplify the
computation by neglecting the dependence of $\psi $ on the $z$
coordinate and therefore the calculations become effectively 2-dimensional.
Accordingly,
vortex states considered here are not subject to three-dimensional
instabilities like corrugation/bending of the vortex
core as the transverse dimension is effectively suppressed.
For this reason often in the following we will describe the
properties of the system as it appears in the x-y plane only.

Eq.\eqref{eq:gpe} is solved either in real time or imaginary time
by using Hamming's predictor-modifier-corrector method 
initiated by a fourth-order
Runge-Kutta-Gill algorithm \cite{Ral60}. 
The spatial mesh spacing and time step are chosen such that,
during the real time evolution, excellent conservation
of the total energy of the system is guaranteed.

We will use in the present work different forms for the 
external potential $V$ appearing in the GP equation \eqref{eq:gpe}.
In particular, we will consider {\it periodic} potentials both in one dimension
and in two dimensions, as described in the following Sections.
These potentials are easily produced in experiments using crossed laser 
fields with appropriate wavelengths (optical lattices, OL).

\section{$^{87}$Rb in 1D optical lattice}

One-dimensional lattice potentials are often used in experiments. 
In the simplest form, like the one used here, they produce 
modulated density patterns characterized by a periodic alternation
of stripes of maxima and minima in the gas density.
Here we consider the form

\begin{equation}
V_{OL}(x)=2V_0\,cos^2(kx)
\label{pot_3}
\end{equation}

The wavevector $k$ determines the lattice constant of the 
periodic potential,
$d=2\pi/k$.
For the cases investigated here we chose $d=60000\,a_0=3.175\,\mu m$.
The sizes of the supercell in the x-y plane are, for most of the calculations done,
$L_x=L_y=44.45\,\mu m$,
corresponding to $14$ stripes in the x direction, although 
for the sake of visibility the plots shown in this paper are
often limited to smaller portions of the supercell.

Along the z-direction we chose the value $L_z=1.429\,\mu m$.
The total number of atoms $N$ in most of our calculations
is chosen so that it corresponds
to an areal density 
$n_a=N/(L_x L_y)=9.92\,\mu m^{-2}$.
In the following, some calculations will be performed using larger
supercell sizes $L_x,L_y$, corresponding to $20$ stripes in the x direction. 
However, we will always take the number of atoms $N$ 
such that the areal density $n_a=N/(L_x L_y)$ is not changed.

The healing length for the uniform system ($V_0=0$) is
$\xi =\hbar/\sqrt{2m\mu}=1.039\,\mu m$, where $\mu = g\rho $ 
is the Rb chemical potential in the absence of the lattice modulation.
Notice that
$L_z$ is of the same order of magnitude of the healing length,
and therefore our system can be considered quasi-2D.
Due to the imposed translational invariance along the $z$ direction,
along which the density is therefore constant, the calculations are effectively 
2-dimensional, involving a spatial discrete mesh only in the x-y plane.

We will consider different values for the potential strength $V_0$
in the following, 
and often the following values $V_0=2,4,7,9 \times
10^{-14}\,Ha$ have been used.

It is customary to express the lattice well depths $V_0$ in terms of the 
so-called recoil energy as $V_0=sE_R$, where

\begin{equation}
E_R=\frac {\hbar ^2 \pi ^2}{2md^2}
\end{equation}
In the present case $E_R=8.600\times 10^{-15}\, $ Ha.
The values of $V_0$ quoted above therefore correspond to $s=2.3,4.6,8.1,10.5$.
For much larger values of $s$ the bosons
are expected to become localized forming rows of
independent quasi-one dimensional superfluids.

In the following we will use a more manageable notation where the 
optical potential strength $V_0$ is expressed in terms of 
an adimensional quantity $\tilde{V}$ such that 
$V_0=\tilde {V}\times 10^{-14}$ Ha $=\tilde {V}\times 3.158\,nK$.
 
\begin{figure}
\centerline{
\includegraphics[width=0.9\linewidth,clip]{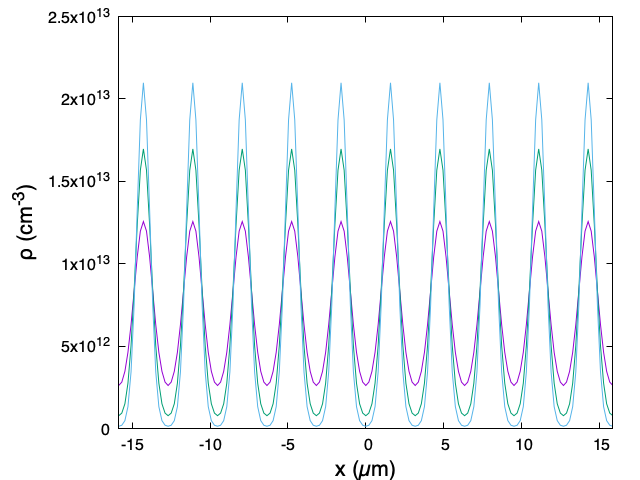}
}
\caption{Density profiles 
(in units of $cm{-3}$)
along the x-axis,
for different 
values of $\tilde{V}$. From the highest to the lowest peak values:
$\tilde{V}=(7,4,2)$.
}
\label{fig1bis}
\end{figure}

In Fig.\ref{fig1bis} we show for some values of $\tilde{V}$ the density along 
the x-axis, the direction of the modulation
showing the periodic alternation of minima and maxima in the density profile.
The values of the density at these
extrema are reported in Table I, together with the calculated chemical potentials.

It can be noticed that
already an intensity $\tilde{V}=4$ the maximum density is ten times
larger of the minimum density. Despite such large
inhomogeneity the system is superfluid.
The superfluid fraction can be computed from the non-classical translational
inertia \cite{sep_joss_rica} as
\begin{equation}
f_s=1-\lim_{v_x\to 0} \frac{\langle \hat{P}_x \rangle}{N m v_x}
\label{sffrac}
\end{equation}
where $\langle \hat{P}_x \rangle = -i\hbar \int \psi ^\ast (\vec r)\partial \psi (\vec r)/\partial x d\vec r$
is the expectation value
of the momentum in the x-direction of the $^{87}$Rb and $N mv_x$ is the total momentum of
the system if all the atoms were moving with the constant velocity $v_x$.

Alternatively, one could estimate the superfluid fraction using Leggett's
formula \cite{leg70}:

\begin{align}
f_s=\frac{L_x^2}{\int _0^{L_x}dx <n(x)> \int _0^{L_x}dx <n(x)>^{-1}}
\label{sffrac_l}
\end{align}
where $<n(x)>$ is the number density of the ground-state configuration 
averaged over the transverse y-z directions.
The above quantity represents a rigorous upper bound to the superfluid fraction, 
and it was shown recently \cite{perez} that
under conditions relevant for most ultracold experiments the two definitions 
\eqref{sffrac} and \eqref{sffrac_l} 
provide surprisingly close values for the superfluid fraction.
We indeed find that the values of $f_s$ from the two definitions above agree with each other 
to within $1\%$, in agreement with the findings in Ref. \onlinecite{perez}.

\begin{figure}
\centerline{
\includegraphics[width=0.90\linewidth,clip]{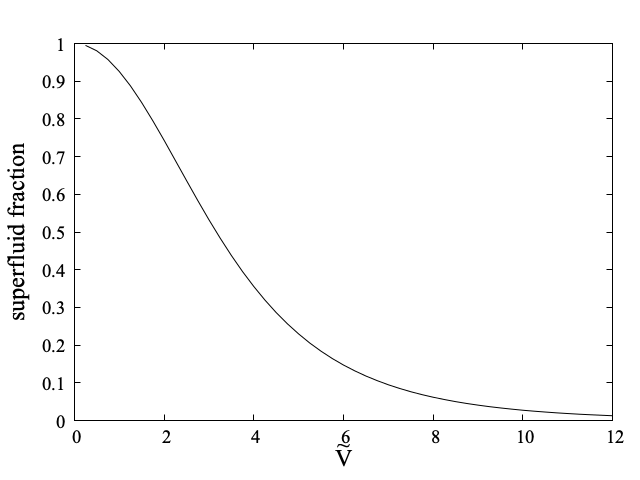}
}
\caption{Superfluid fraction in the 1D lattice as a function of $\tilde{V}$
in the direction of the modulation.
}
\label{sf_1d}
\end{figure}

The dependence on $\tilde{V}$ of the superfluid fraction (in the direction parallel to the
modulation) is shown in Fig.\eqref{sf_1d}
and the values of
$f_s$ for some values of $\tilde{V}$ are given in Table I. We notice
that the suppression of the superfluid fraction in the
direction of the modulation is quite substantial. In the
direction perpendicular to the modulation, i.e. in the y direction, 
the superfluid fraction is unity in all cases. Therefore we have a strongly
anisotropic superfluid.

\begin{table*}
\begin{tabular}{cccccccccc}
\hline
& $\tilde{V} $  & $\rho _{min} $ $ (cm^{-3}) $  & $\rho _{max} $ $ (cm^{-3}) $  & $\mu $ ($ nK $) &   $f_s$ \\
\hline
 & $0$ & $6.943\times 10^{12}$  & $6.943\times 10^{12}$  & $2.584 $ & $1$  \\
 & $2$  & $2.609\times 10^{12}$  & $1.256\times 10^{13}$ & $8.062 $ &  $0.744 $  \\
 & $4$  & $0.777\times 10^{12}$  & $1.696\times 10^{13}$ & $11.923 $&  $0.355 $  \\
 & $7$  & $0.136\times 10^{12}$  & $2.098\times 10^{13}$ & $15.839 $&  $0.095 $  \\
\hline
 \end{tabular}
  \caption{ $\tilde{V}$ is the amplitude of the periodic potential; 
$\rho _{min} $ and $\rho _{max} $ are the density values at the bottom 
and top of the modulated density;
$\mu $ is the chemical potential and $f_s$ is
the superfluid fraction in the x-direction computed from the non-classical translational inertia.  \\
 \label{table3}
}
\end{table*}

\subsection{Single vortex properties}

A linear, singly quantized vortex excitation in
the z direction,
with the core in the position $(x_v,y_v)$,
can be generated by the ``phase imprinting'',
i.e. we compute the lowest energy state
obtained by starting the imaginary time evolution from
the initial wave function
\begin{equation}
\psi_v(\mathbf{r})=\rho_0^{1/2}(\mathbf{r})\, \left[ {(x-x_v)+i (y-y_v) 
\over \sqrt{(x-x_v)^2+(y-y_v)^2}}  \right]
\label{imprint}
\end{equation}
where $\rho_0(\mathbf{r})$ is the
ground-state density of the vortex-free system. 
This wave function has unit
circulation in the x-y plane and it is orthogonal to the ground state. Such
properties are maintained during the imaginary time evolution until the lowest
energy state with such properties is reached.
During this evolution, the vortex position and core structure 
change to provide at convergence the lowest energy
configuration for a vortex with unit of circulation 
$h/m$ in clockwise direction. The sign of the circulation can be changed by
changing the sign of the complex $i$ in Eq.(\ref{imprint}).
We have verified that it is
possible to study the dynamics of a vortex also in real time when it is started
from a non-stationary position. In fact in this case we find that the evolution in
imaginary time has a rapid transient during which the imprinted phase and
modulus of Eq.\eqref{imprint} are modified reflecting the presence of the external potential.
During this stage the initial vortex core position is essentially unaffected and it
is only for much larger imaginary times that the vortex core position migrates
to the nearest equilibrium position. The protocol we follow for real time study
of a vortex is therefore to perform a short imaginary time evolution and after to evolve
the system in real time.
 
The flow field of a linear vortex has a long-range character, 
$\sim 1/r$, where $r$ is the distance from the position of the vortex core.
We have imposed antiperiodic boundary conditions \cite{Pi07} in order 
that the condition of no flow across the boundary of 
the computational cell is satisfied \cite{Anc21}.
This is equivalent 
to sum over the phases of an infinite array of vortex-antivortex, 
i.e. a vortex of opposite chirality is present in each 
nearest neighbor cell of the computation cell \cite{sadd}. 
Equation \eqref{imprint} can be 
easily generalized to accommodate a vortex array made of an arbitrary number
of vortices and/or antivortices \cite{Anc18}. 
The case of a pair of vortices with opposite circulation
(vortex dipole) will be considered in the following. In the case of a vortex dipole
we impose however the usual periodic boundary conditions since the flow fields of
the two vortices tend to cancel at the cell boundaries.

\begin{figure}
\centerline{
\includegraphics[width=0.8\linewidth,clip]{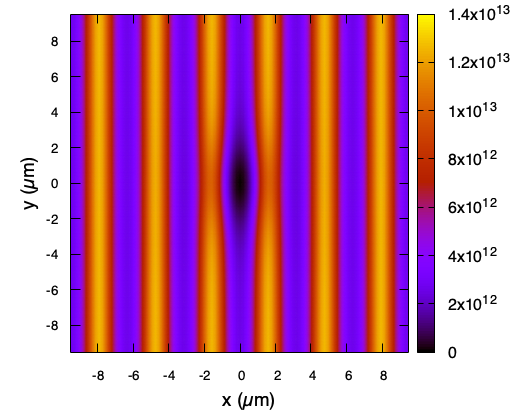}
}
\caption{ 
Structure of a singly-quantized vortex at x=y=0, one of the stable sites
at the density minimum, for the 
case $\tilde{V}=2$.
The density is in units of $cm^{-3}$.
}
\label{fig3v}
\end{figure}

In general, we find that the stable vortex positions are at the
sites of minimum density, i.e. at the locus of maxima of the
external potential (see also the following Sections, where different
types of optical lattices are employed), in agreement with earlier observations \cite{duine}. 
In the case of the present one-dimensional lattice potential 
the stable positions correspond to the set
of lines in the y direction (or better planes if we consider
also the z direction) where the density is minimum.
Therefore these lines are at the bottom of a kind of channels
along which a vortex can freely move with no change of energy.

\begin{figure}
\centerline{
\includegraphics[width=0.7\linewidth,clip]{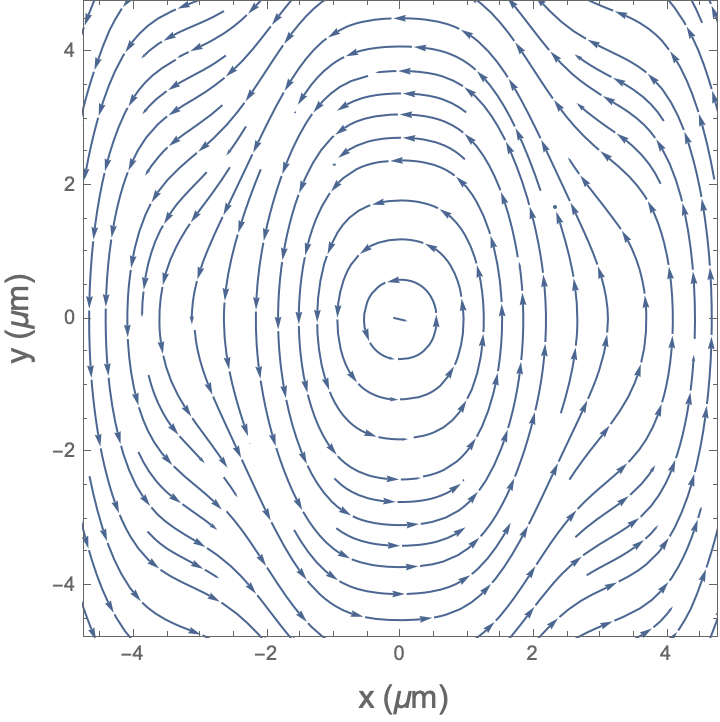}
}
\caption{Streamlines for a vortex in the one-dimensional lattice, for the case $\tilde{V}=2$.
The x and y axis show coordinates in $\mu $m.
}
\label{fig_stream_3}
\end{figure}

The equilibrium vortex structure at one of such sites 
is shown in Fig.\ref{fig3v} for the case $\tilde{V}=2$.
The streamlines for the velocity field 
(lines that are tangential to the instantaneous velocity direction)
$\vec v=\hbar \nabla \phi /m$ ($\phi  $ being the phase of the wavefunction)
are shown in Fig.\ref{fig_stream_3}.

\begin{figure}
\centerline{
\includegraphics[width=0.7\linewidth,clip]{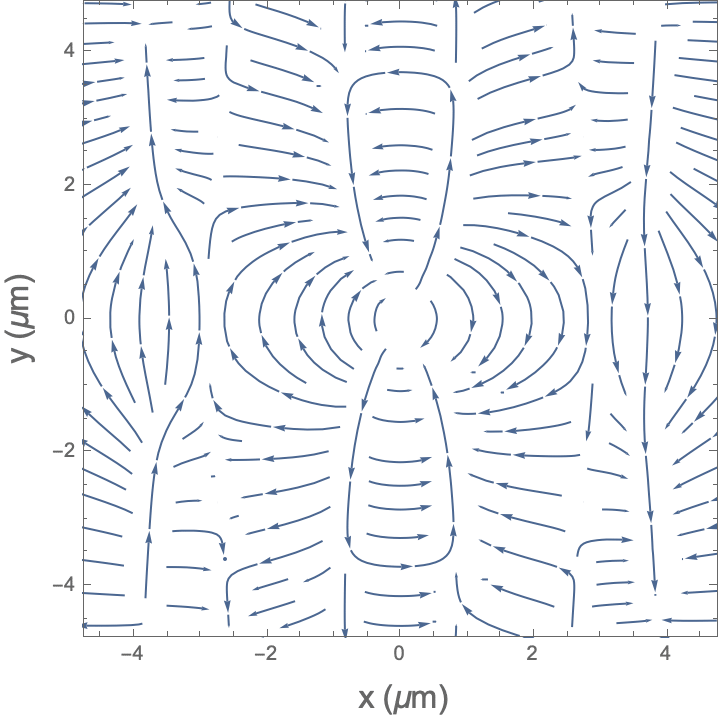}
}
\caption{Streamlines of $\vec \nabla \tilde {\phi}$ 
for a vortex in the one-dimensional lattice, for the case $\tilde{V}=2$.
The x and y axis show coordinates in $\mu $m.
}
\label{fig_stream_4}
\end{figure}

Notice the strong deformations with respect to the circular patterns expected from 
a vortex in a homogeneous system.
In order to display these deformations more
clearly in Fig.\ref{fig_stream_4} 
we also show the streamlines of the modified phase 
${\tilde{\phi}}$ as defined in Eq.\eqref{psi_tilde} further on.
${\tilde{\phi}}$
represents the deviation of the phase from the imprinted phase in Eq.\eqref{imprint}.

Besides the stable positions at the bottom of the low density channels,
there are also metastable
positions at the density maxima,
with higher energy than the equilibrium one.
The associated energy barrier for vortex migration
from one channel to the
neighboring one depends on the amplitude of the modulation.

In Table \ref{table4}
the vortex excitation energy and the energy barrier,
both per particle and per unit
length of the vortex, as well as the angular momentum are
reported for selected values of $\tilde{V}$. The vortex excitation energy
$\Delta E_v$ is finite because the computation box is finite (in an
infinite system $\Delta E_v$ would diverge 
with the size of the system in a logarithmic way).
$\Delta E_v$ in the modulated system is smaller than in the uniform
system and it decreases for increasing $\tilde{V}$ because the vortex
core is located in a region where the density decreases as $\tilde{V}$
increases. The energy barrier is a strongly increasing
function of $\tilde{V}$ and this reflects the strong variation of
the ratio $\rho _{max}/\rho _{min}$. We notice also 
the reduced value of the angular
momentum and this reflects the reduced superfluid fraction in
presence of the modulation. 
In the homogeneous state ($\tilde{V}=0$)
the angular momentum deviates from the theoretical value $N\hbar $
due to the boundary effect of the computation box.

\begin{table*}
\begin{tabular}{cccccccccccc}
\hline
& $\tilde{V}\,\,\,\, $ & $\Delta E_v/N$ ($ nK/\mu m $) & $\,\,\,\,(E_B-E_H)/N$ ($ nK/\mu m $) & $\,\,\,\,<\hat{\mathcal{L}}_z>$ ($N\hbar $) \\
\hline
 & $0$ & $ 0.022 $ & - & $0.925$   \\
 & $2$  & $0.018$ & $0.0026\,$ & $\,0.788$ \\
 & $4$  & $0.011$ & $ 0.0062\,$ & $\,0.492$ \\
 & $7$  & $0.005$ & $ 0.0109\,$ & $\,0.198$ \\
\hline
 \end{tabular}
  \caption{ 
$\Delta E_v=(E_H-E_0)/L_z$ is the energy (per unit length) cost to create a vortex in the minimum density 
sites, $E_0$ being the energy value (per atom) in the absence of the vortex;
$(E_B-E_H)/L_z$ is the energy barrier (per unit length) to move a vortex across the maximum density ridge;
$<\hat {\mathcal{L}}_z>$ is the angular momentum along the z-axis in units of $N\hbar $.
\\
 \label{table4}
}
\end{table*}

\subsection{Vortex dipole properties}

Due to the presence of particular stable sites for a vortex 
caused by the presence of the spatial periodicity imposed by the optical 
lattice potential, a number of properties are expected to
differ from those in a (nearly homogeneous) superfluid. 

In classical hydrodynamics of incompressible fluids,
a vortex dipole is a stable entity that moves with a constant
velocity that is perpendicular to the plane defined by the
vortex axis and the vector $\vec l$ joining the vortex and the
antivortex core positions 
and inversely proportional to the distance $l$ between
them. The same behavior holds in superfluid
systems \cite{donnelly,roberts} when
$l$ is much larger than the healing length, and
the vortex dipole propagates with a
constant velocity $v_d = \hbar /(ml)$.
For example, in the absence of any modulation,
a vortex-antivortex pair separated by a distance 
equal to $l=4d=12.7\,\mu m$
is found to translate with a constant velocity $0.058\pm 0.002$ $\mu m/
ms$ (the error bar being estimated from the fit to the calculated data),
to be compared with the hydrodynamical theoretical value
$v_d=0.057$ $\mu m/ms$.

We have studied the real-time evolution of a vortex dipole imprinted
in the superfluid in a 1D optical lattice described above.
The position of the vortex
core is located at each time step by carefully scanning the spatial mesh to find 
the minimum in the density corresponding to the vortex core.
In particular, we will consider two different initial
states for the vortex-antivortex pair, i.e.
(i) the vortices are located in different minimum energy channels, at initial positions $x=\pm 2d$,
where $d$ is the lattice constant;
(ii) the vortices are located in the same channel (at $x=0$), at initial positions $y=\pm 2d$.

\begin{figure}
\centerline{
\includegraphics[width=0.8\linewidth,clip]{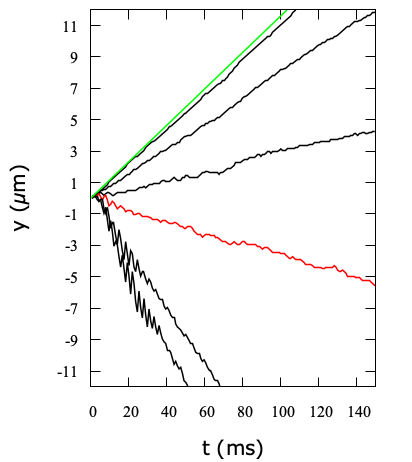}
}
\caption{ 
Trajectories along $y$ of a dipole with the vortices in
two different channels.
The vortices are initially 
set at positions $x=\pm 2d$ and with the same $y$. 
The $y$ coordinates of the two
vortices of the dipole remain equal at all times within the grid of the
computation.
From top to bottom: $\tilde{V}=0,1,2,3,4,6,7$.
}
\label{fig_dipolex_stripe_hor}
\end{figure}

In the case (i) we find that the vortex pair move with a roughly constant
velocity, the two vortices remaining in the same initial channel.
The translation velocity decreases as the intensity
$\tilde{V}$ increases until between $4$ and $6$
the velocity changes
sign, i.e. it is in the opposite direction of motion compared to
that of the homogeneous superfluid.
This is shown in Fig.\ref{fig_dipolex_stripe_hor}.
This is a novel effect not
seen before. 
Superimposed on the translation is a weak oscillation for small 
values of $\tilde{V}$ but the oscillation becomes very intense 
for large values of $\tilde{V}$, for the largest intensity the motion 
periodically changes direction. The period of this oscillation 
becomes shorter for increasing $\tilde{V}$. For $\tilde{V}=7$ the period 
is about $1.9\, ms^{-1}$. We will discuss more quantitatively 
the origin of some of these effects in the following.

This behavior of the dynamics of the vortex dipole in distinct
channels appears to be quite general: a) we have verified this when
the vortex and the anti vortex lie in farther apart channels,
b) when the vortex and the anti vortex start with a different
$y$ coordinate they rapidly synchronize their motion at a common
$y$ coordinate, i.e. they minimize their distance remaining in
the initial channels.

\begin{figure}
\centerline{
\includegraphics[width=0.8\linewidth,clip]{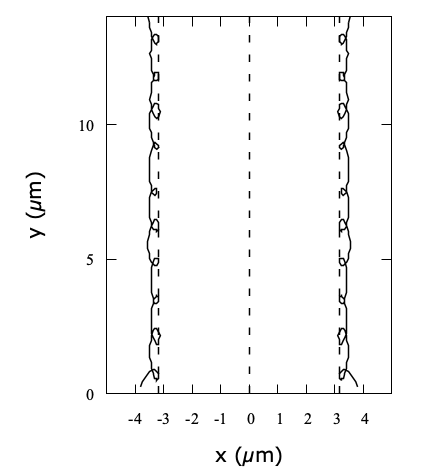}
}
\caption{Dynamics of a vortex dipole when the initial vortex positions are slightly 
displaced with respect to the channel axis. The three vertical lines show the 
locii of minimum density. The simulation refers to the case $\tilde{V}=2$.
}
\label{fig_dipole_core_stripe}
\end{figure}

A rather peculiar behavior occurs when the initial positions
of the two
vortices in different channels do not coincide with the stable positions at the
density minima,
but are rather slightly displaced with respect to it, in
opposite directions.
During the ensuing dynamics, the vortices are found to move
along the channels
with complex trajectories shown in Fig.\ref{fig_dipole_core_stripe}. Such
trajectories are the result of a uniform translation along the
$y$ direction and of two oscillatory motions (occurring with the
same
frequency) along x and y directions. It should be stressed
that this oscillatory motion perpendicular to the direction of
translation is not an oscillatory motion around the position
of minimum energy but it is rather a one-side oscillation: 
if the vortex starts to
the right of the minimum at all times it remains on this side
of the minimum and it never crosses to the other side. The
same holds it it starts to the left. Therefore this dynamics
is quite distinct from that of a massive particle moving
around a minimum energy position.

We note that
small-amplitude oscillations superimposed to vortex trajectories
has been found in a two-component Bose mixture \cite{richaud},
within the so-called massive point-vortex model,
for a single vortex with a rigid circular boundary, where a massless vortex can 
only precess uniformly. 
These phenomenology and the one unveiled by our calculations
certainly share some common physical properties.
The model of Ref. \onlinecite{richaud} is based on a Lagrangian formulation 
for point-vortices, and has been applied so far to homogeneous systems \cite{richaud,richaud1}.
It might be of great interest to extend such model to the case of non-homogeneous 
systems like the one described here.

\begin{figure}
\centerline{
\includegraphics[width=0.8\linewidth,clip]{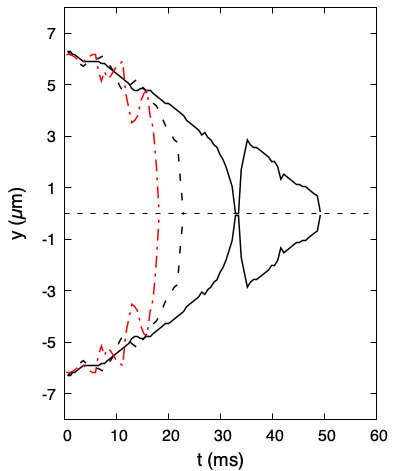}
}
\caption{ 
Trajectories along y of the two vortices of a dipole
started in the same channel.
The vortices are initially 
set at positions $y=\pm 2d$. From left to right $\tilde{V}=7,4,2$. 
}
\label{fig_dipolex_stripe_vert}
\end{figure}

In the case (ii), vortex and anti-vortex started in the same channel, for 
$\tilde{V}=4,7$ the two vortices moves towards one another,
remaining in the same channel, until they
annihilate at the origin, as shown in Fig.\ref{fig_dipolex_stripe_vert}.
However, in the case of a less modulated structure ($\tilde{V}=2$), they remain
initially
in the same channel and move towards one another until they nearly touch
at time $t=32\,ms$ (still being in the density minimum same channel):
at this point, they appear to repel each other and jump to the next channel,
annihilating there
at a later time.
We notice that a similar behavior has been observed in
experiments \cite{neely}, where
vortices in BEC are found to
approach each other so closely that they appear to
coalesce, but then move away from each other after the close encounter.
This combination of jump and approach of the
vortex pair two vortices of the dipole has some similarity with what we find in
the case
of 2D modulations as described in the next subsection.

From the equation for the velocity of a
vortex within the GP equation we can understand at a
qualitative level some aspects of our results. The flow field of one
vortex
on the other is in the transverse direction of the channel,
i.e. in the region where the
density of the superfluid is increasing. This variation
of density gives rise to a component of the velocity of the
vortex
in the longitudinal direction along the bottom of the
channel. Following Ref. \onlinecite{simula} (see also Ref. \onlinecite{sep_joss_rica}) 
one may write the vortex wave function $\psi _v (\vec r,t)$
with the core located at position $\vec r_0$ at time $t$ as
\begin{equation}
\psi_v (\vec r,t)=(x+iy-x_0-iy_0)\tilde{\rho }e^{i\tilde{\phi}}
\label{psi_tilde}
\end{equation}
(here $\vec r_0=(x_0,y_0)$ is expressed in complex notation as $z_0=x_0+iy_0$),
where the factor $z=x+iy-x_0-iy_0$ represents the ideal gas vortex wave function 
and $\tilde{\rho }(\vec r,t)$ and $\tilde{\phi }(\vec r,t)$ represent the deviations 
of the modulus and of the phase from this ideal gas form (see Ref. \onlinecite{simula} 
for additional details). Several effects contribute
to the deviation of the vortex wave 
function from the ideal gas form, i.e. $\tilde{\rho }(\vec r,t)$ and $\tilde{\phi }(\vec r,t)$ 
have the contribution due to the presence of other vortices, 
to the effect of an external potential and, finally, to the 
deviation of $\psi_v (\vec r,t)$ from the ideal gas form caused by the inter-atomic interactions.
On the basis of the Gross-Pitaevskii
equation here adopted the velocity of the vortex is \cite{simula}
\begin{equation}
\vec v = \frac {\hbar }{m}(\nabla \tilde {\phi }-\hat {\kappa }\times \nabla log \tilde {\rho })
\label{eq:tilde}
\end{equation}
where the vector $\hat {\kappa } $ is the unit vector in the direction of the circulation,
i.e. in the z direction in the present case. As soon as
the vortex position moves out of the bottom of the channel the
second contribution to the velocity in Eq.\eqref{eq:tilde} becomes non-zero due to 
the increasing density and its gradient is in the x direction.
The vector product of these two vectors is in the y direction and
points toward the other vortex. Therefore the velocity of one
vortex of the pair has one longitudinal contribution from the
external potential and a transverse contribution from the gradient
of the phase due to the other vortex of the pair. This
transverse term might dominate if the amplitude of the external
potential is weak enough so that the vortex pair could move across
channels but in any case due to the longitudinal contribution the two vortices
move also against each other until
they annihilate. The longitudinal contribution to the vortex
velocity might dominate for large amplitude of the external potential
so that the two vortices remain in the starting channel until
annihilation.
Eq.\eqref{eq:tilde} for the velocity of the vortex core also explain why we see one-side
oscillations of the vortex: the gradient of the local density tends to bring the vortex core
toward the minimum density
position but at this minimum this contribution to the velocity vanishes.

It should be noticed at this point that Eq.\eqref{eq:tilde} is 
not a self-contained equation for the dynamics of a vortex 
because the tilded quantities are just a different way 
to represent the solution of the time-dependent GP equation.
As such, Eq.\eqref{eq:tilde} can help to develop 
qualitative arguments as exposed above,
but it cannot predict the dynamics unless approximations are introduced.

On the basis of Eq.\eqref{eq:tilde} for the velocity of a vortex we can also understand why
the translation velocity of the vortex dipole lying in two different channels is
reduced with respect to the homogeneous case. In fact from
Fig.\ref{fig_stream_4} one can see that the direction of ${\nabla \tilde{\phi}}$ along the line
corresponding to $y=0$ has the opposite direction of the one of the
homogeneous case, i.e. it points downward for $x>0$ and upward for $x<0$ and
not vice versa as should be for a positive circulation vortex in a homogeneous
superfluid. This means that the induced velocity due to the other vortex of the
dipole is reduced compared with the value for the homogeneous system. Why
the translation velocity even changes direction at large modulations and a
longitudinal oscillation is present, however, cannot explained by these simple
considerations.

\subsection{Single vortex in a trap}

An important question is if the oscillations in the
dynamics of a vortex pair displayed in Fig.\ref{fig_dipole_core_stripe} are a consequence
of the mutual coupling between vortex-antivortex or are rather
a property of the single vortex.
We therefore considered the dynamics of a single vortex
initially
imprinted in a position slightly displaced from a position of
minimum energy by the same amount as in the
dipole simulation just described. However boundary effects
mask the genuine dynamics in the case of an 
extended system with anti-periodic boundary conditions. To
avoid this we used instead a finite system confined
in the x-y plane by an additional circular "box" potential of the form

\begin{equation}
V_{box}(x,y)=U_0\left[1-\frac {1}{e^{(R-R_0)/\sigma }+1}\right]
\end{equation}
where $R=\sqrt{x^2+y^2}$, $R_0$ is the chosen value for the radius of the 
circular trap.
We chose the following values:
$R_0=14.3\,\mu m$, $U_0=60\,nK$, and $\sigma =0.212\,\mu m$.
With such choice, the potential $V_{box}$ becomes different from 
zero as $R$ approaches $R_0$ and rapidly becomes very repulsive reaching 
the value $U_0$. As a result, the system is essentially unaffected in the 
inner region by this potential, where the system density is very close to that of the 
extended system, and goes exponentially to zero for radial distances 
from the center larger that $R_0$.
We chose the number of atoms so that the density pattern in the interior of the 
circular trap is very similar to that of the extended system, for a given value of $\tilde{V}$.

We perform the computation with the external potential $V(\vec r)= V_{OL}(x,y)
+V_{box}(x,y)$ and $V_{OL}$ is such that the center of the trap is a maximum of the
OL, 
i.e. the center of the trap is an equilibrium position for the
vortex core. We find that the vortex remains immobile if
initially it is located at or very near 
the center of the trap, in the low density channel passing through it.
If the initial position of the vortex is off-center along the y direction
by a larger amount, however, the vortex is found to migrate 
towards the edge of the trap and disappears once it approaches the trap edge.

\begin{figure}
\centerline{
\includegraphics[width=0.8\linewidth,clip]{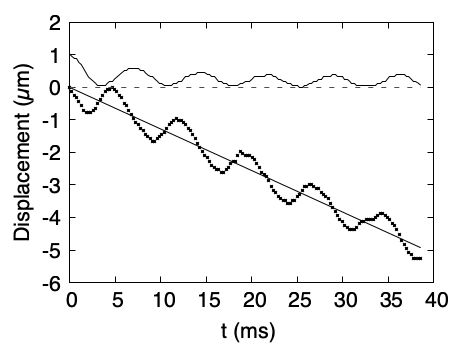}
}
\caption{Upper curve: displacement $x(t)$ of the vortex core perpendicular
to the channel axis; lower curve: displacement $y(t)$ of the vortex core parallel
to the channel axis. The dotted line shows a least-square fit to $y(t)$,
The simulation refers to the case $\tilde{V}=2$.
}
\label{fig_core_motion}
\end{figure}

When the initial position is at $y=0$ but slightly off center 
along the x-direction, i.e. away
from an equilibrium position, then the vortex moves toward the
border of the trap and the radial motion is the superposition
of a uniform motion in the radial direction and of a longitudinal and transverse
oscillatory motion, similarly to the case of a vortex dipole
shown in Fig.\ref{fig_dipole_core_stripe}. As in the case of the vortex dipole the
oscillatory motion is one-sided: the vortex never 
crosses the line of minimum density
but it remains on the same side of the channel from which it
started.
This is shown in Fig.\ref{fig_core_motion}.

\begin{figure}
\centerline{
\includegraphics[width=0.8\linewidth,clip]{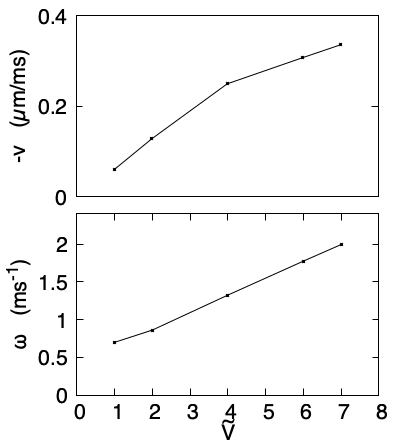}
}
\caption{Upper panel: average translation velocity of a single vortex
starting from a slightly displaced x-position from the center of the trap;
Lower panel: calculated frequency of the oscillations occurring 
during the vortex motion along the channel.
}
\label{fig_dipolex_stripe}
\end{figure}

We have estimated, from the trajectories $x(t)$ and $y(t)$, 
the frequency characterizing these oscillations, which is reported in 
Fig.\ref{fig_dipolex_stripe} as a function of $\tilde{V}$, together with the
average translation velocity of the vortex along the channel.

\bigskip
\bigskip

We finally note that the frequencies found for the oscillations
of a single vortex, as discussed above, agree with the ones found for the
case of a vortex dipole propagating in parallel channels.

We have checked the effect of changing the density of the system on the
oscillation frequency in Fig.\ref{fig_dipolex_stripe}. We thus studied the dynamics of a 
vortex slightly displaced from the equilibrium position
in the transverse direction x, as described before, but when the 
total number of atoms in the system is doubled (halved). 
We find that both the amplitude and frequency of the transverse oscillations
are unaffected by the increased (decreased) density. 
Also the frequency of the back-and-forth oscillations along the 
channel axis is not changed, whereas its amplitude is reduced
as the density increases.

A natural question is if these oscillations might be due 
to the coupling of the vortex motion with the 
phonon excitations in the system which can be created by the vortex
motion through the system. However, this is not the case. 
In fact we have computed, by using the Bogoliubov-de Gennes approach,
the dispersion relation perpendicular and parallel to the channel 
direction. The phonons propagating in a direction perpendicular to the 
channels, having a dispersion relation which flattens at the Brillouin zone
boundary, are those with the highest density of states. Their frequencies
are found to be \cite{anc_next} in the 
range $0.12<\omega <0.4 $ ms$^{-1}$ (corresponding to the range $7>\tilde{V}>2$), i.e. 
much smaller than the observed vortex oscillation 
frequencies. 

\section{$^{87}$Rb in 2D optical lattice}

\subsection{Square lattice}

The external potential $V(\vec r)$ acting on the Rb atoms is 
taken in the form of a periodic potential with 
square symmetry, i.e. $V({\bf r})=V_{OL}(x,y)$, where

\begin{equation}
V_{OL}(x,y)=V_0[cos^2(\kappa x)+cos^2(\kappa y)]
\label{pot_2}
\end{equation}

As done in the previous Sections, 
the optical potential strength $V_0$ is expressed in the following in terms of 
an adimensional quantity $\tilde{V}$ such that 
$V_0=\tilde {V}\times 10^{-14}$ Ha $=\tilde {V}\times 3.158\,nK$.

We first computed the ground-state in the 
presence of the periodic potential $V_{OL}$, for different values of the 
amplitude $\tilde{V}$. 

We will use in the following a shorthand notation where 
the site of maximum density (corresponding to the minimum of the
lattice potential) is called 'top' (T), the site with 
minimum density (corresponding to the maximum of the lattice potential) 
is called 'hollow' (H), and the saddle point between
two adjacent top sites is called 'bridge' (B).

As in the case of the 1D potential discussed in Section III,
the system is superfluid, with a superfluid fraction $f_s$
which depends on the amplitude of the lattice modulation $V_0$.
The relevant densities at the various lattice sites, and the
calculated superfluid fraction are reported in Table \ref{table1}.

\begin{table*}
\begin{tabular}{ccccccccccc}
\hline
& $\tilde{V} $ & $\rho _{H} $ $ (cm^{-3}) $  & $\rho _{T} $ $ (cm^{-3}) $  & $\rho _{B}$ $ (cm^{-3}) $ & $\mu (nK)$ &  $f_s$ \\
\hline
 & $0$ & $6.943\times 10^{12}$  & $6.943\times 10^{12}$ & $6.943\times 10^{12}$ & $2.584$  & $1$  \\
 & $2$  & $2.797\times 10^{12}$  & $1.338\times 10^{13}$ & $6.462\times 10^{12}$  & $8.48$& $0.929 $  \\
 & $4$  & $0.9412\times 10^{12}$  & $2.079\times 10^{13}$ & $5.295\times 10^{12}$  & $13.57$& $0.768 $  \\
 & $7$  & $0.1668\times 10^{12}$  & $3.151\times 10^{13}$ & $3.329\times 10^{12}$  & $19.97$& $0.513 $  \\
 & $9$  & $0.0537\times 10^{12}$  & $3.793\times 10^{13}$ & $2.288\times 10^{12}$  & $23.59$& $0.375 $  \\
\hline
 \end{tabular}
  \caption{ $\tilde{V}$ is the amplitude of the periodic potential; 
The density values at the Hollow, Top and Bridge sites are shown; 
$\mu $ is the chemical potential; $f_s$ is
the superfluid fraction computed from the non-classical translational inertia. \\
 \label{table1}
}
\end{table*}

\begin{table*}
\begin{tabular}{lllllllllll}
\hline
& $\tilde{V} $ & $\Delta E_v/N$ ($ nK/\mu m $)  & $(E_B-E_H)/N$ ($ nK/\mu m $) & $(E_T-E_H)/N$ ($ nK/\mu m $) & $<\hat{\mathcal{L}}_z>$ ($N\hbar $) \\
\hline
 & $0$ &  $0.0220$ &  -  & - & $\,0.925$ $(0.927^\ast )$ \\
 & $2$  & $0.0199$ &  $ 0.00049\,$ & $0.00210$ & $\,0.858$\\
 & $4$  & $0.0163$ &  $ 0.00046\,$ & $0.00468$ & $\,0.708$\\
 & $7$  & $0.0108$ &  $ 0.00026\,$ & $0.00886$ & $\,0.486$ \\
 & $9$  & $0.0078$ &  $ 0.00019\,$ & $0.01147$ $(unstable)$ & $\,0.356$\\
\hline
 \end{tabular}
  \caption{ 
$\Delta E_v=(E_H-E_0)/L_z$ is the energy (per unit length) cost to create a vortex in the minimum density 
(hollow) sites, $E_0$ being the energy value (per atom) in the absence of the vortex;
$<\hat{\mathcal{L}}_z>$ is the angular momentum along the z-axis in units of $N\hbar $
(the value at $\tilde{V}=0$ denoted with an asterisk has been computed with a bigger cell, 
with a surface area in the x-y plane 4 times larger the one used for all the calculations;
$(E_B-E_H)/L_z$ and $(E_T-E_H)/L_z$ are the energy barriers (per unit length) to move 
a vortex across the bridge site and the top site, respectively. 
\\
 \label{table2}
}
\end{table*}

\subsection{Vortices}

\begin{figure}
\centerline{
\includegraphics[width=1.0\linewidth,clip]{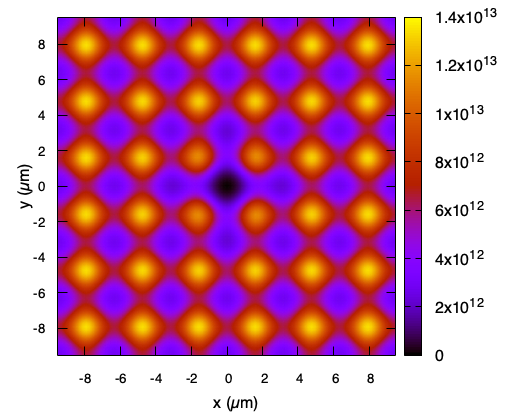}
}
\caption{ 
Structure of a singly-quantized vortex in the stable hollow site, for the 
case $\tilde{V}=2$. The vortex core is at the origin $(x=y=0)$.
The density is in units of $cm^{-3}$.
}
\label{fig3}
\end{figure}

We imprint a single vortex using the procedure described before.
If the vortex initial position is in a generic point
of the unit cell of the modulating potential, during the
imaginary time evolution the vortex core moves to the closest
minimum density position.
We find that stable vortex positions are at the low density sites corresponding to the 
maxima of the optical potential (hollow site), as discussed in the following.
In the case of a square lattice these
equilibrium positions have the same square symmetry.
The vortex structure at the hollow site 
is shown in Fig.\ref{fig3} for the case $\tilde{V}=2$.

\begin{figure}
\centerline{
\includegraphics[width=1.0\linewidth,clip]{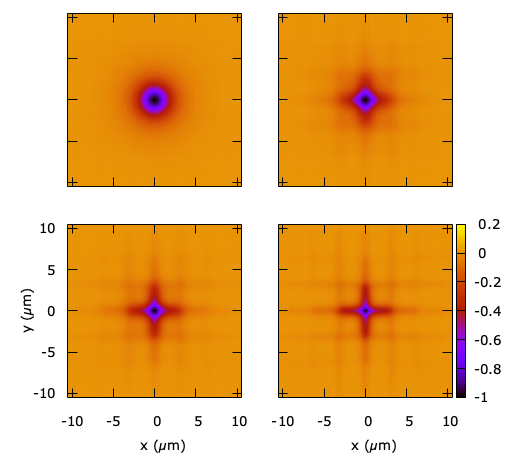}
}
\caption{Relative density difference $(\rho _{vort}-\rho _{novort})/\rho _{novort}$ 
showing the 
minimum energy vortex structure (hollow site)
for $\tilde{V}=0,2,4,7$.
}
\label{fig4}
\end{figure}

We show in Figs.\ref{fig4} the equilibrium vortex structure 
for different values of the potential depth $\tilde{V}$, including the value $\tilde{V}=0$.
In order to improve the visualization of the vortex core structure, 
we show in the figures the density differences
with respect to the configuration without vortex.

\begin{figure}
\centerline{
\includegraphics[width=0.8\linewidth,clip]{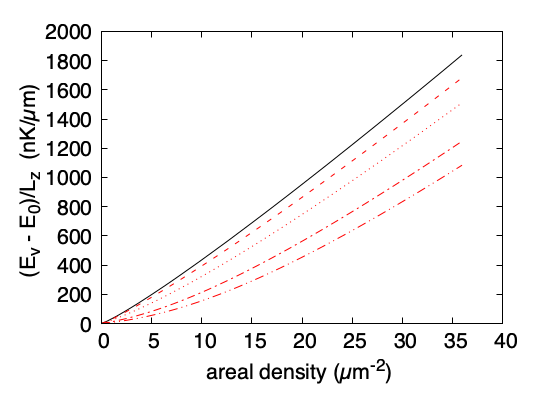}
}
\caption{Vortex total energy 
as a function of the areal density of Rb atom.
From top to bottom:
$\tilde{V}=0,2,4,7,9$. 
The density $n_a=9.92\,\mu m^{-2}$ is the 
one used in most of the calculations discussed in the paper.
}
\label{fig3_ene}
\end{figure}

We show in
Fig.\ref{fig3_ene} the calculated vortex excitation energy per
unit length $\Delta E_v
\equiv (E_v-E_0)/L_z$ as a function of density 
of Rb atoms.

Notice the decrease of the vortex energy, for a given $N$, with the 
modulation amplitude of the lattice potential.
This is a consequence 
of the decreasing density in the region of the vortex core
as $\tilde{V}$ increases.

\begin{figure}
\centerline{
\includegraphics[width=0.7\linewidth,clip]{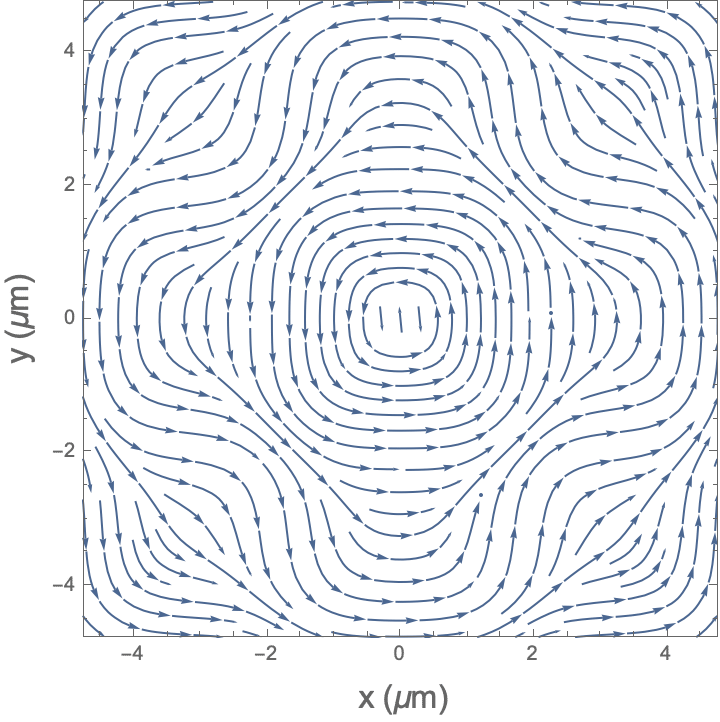}
}
\caption{Streamlines for a vortex in the hollow site, for the case $\tilde{V}=7$.
The x and y axis show coordinates in $\mu $m.
}
\label{fig_stream_2}
\end{figure}

The streamlines around the vortex core are shown in 
Fig.\ref{fig_stream_2} for $\tilde{V}=7$.

\begin{figure}
\centerline{
\includegraphics[width=1.0\linewidth,clip]{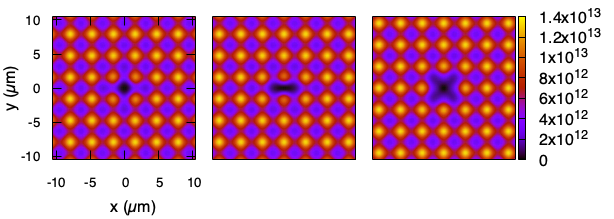}
}
\caption{ Structure of a singly-quantized vortex, for the 
case $\tilde{V}=2$ (only a portion of the density in the x-y plane is shown).
From left to right:
(i) hollow (most stable) site; (ii) bridge site; (iii) top site.
The density is in units of $cm^{-3}$.
}
\label{fig6b}
\end{figure}

Besides the stable position at the hollow site,
we have found that the
saddle point between two adjacent density minima ({\it bridge} site in the following)
is a metastable equilibrium position for a vortex, with slightly higher energy than
the equilibrium one.
A third metastable position for the vortex is found,
at least
for not too large value of $\tilde{V}$, at the high-density sites ({\it top} site in the following),
corresponding to the minima of the optical potential. 
We show the vortex structures for these configurations in 
Fig.\ref{fig6b}, together with the stable hollow configuration.

In a uniform superfluid the energy of a vortex does not depend
on the position of its core so that it is free to move under
the influence of a perturbation. Our results show that in
presence of a 2D optical lattice the vortex energy
becomes a periodic function of position.
The associated energy barriers for vortex migration 
along the lattice are reported in Table \ref{table2}.
These barriers will play an important role, as discussed 
in the following, in the dynamics of 
vortices across the lattice.

\bigskip

\subsection{Vortex dipole properties}

Following the protocol described in Sect.III.B we imprint a vortex dipole in the
2D modulated superfluid and follow its real time dynamics.
Instead of rigidly translating
as in an homogeneous superfluid phase, in the presence of
spatial modulation the vortex and antivortex 
approach each other by a series of jumps from one site
to another moving
mostly across the saddle positions until
they annihilate in a very short time and their energy is released in the 
form of density wave excitations.

\begin{figure}
\centerline{
\includegraphics[width=0.9\linewidth,clip]{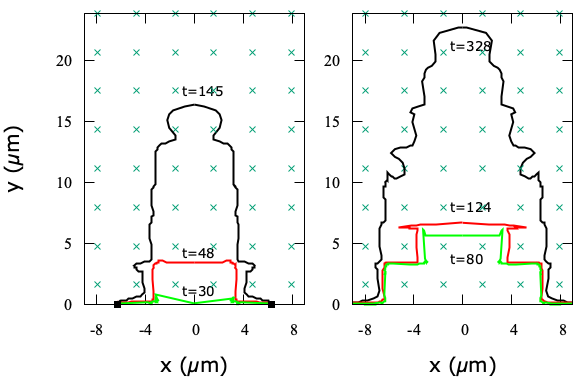}
}
\caption{Trajectories of the two vortices of the dipole in the x-y plane ending in
vortex dipole annihilation. 
Left panel: the two vortices are initially placed at $x=\pm 2d=\pm 6.35\,\mu m$.
From top to bottom:
$\tilde{V}=2,4,7$. The label on each trajectory gives the observed annihilation
time in milliseconds. The crosses show the positions of the T sites where the Rb density is maximum.
Right panel: the two vortices are initially placed at $x=\pm 3d=\pm 9.53\,\mu m$.
}
\label{fig_core_1}
\end{figure}

The path followed during the annihilation process depends on the 
amplitude $\tilde{V}$ of the optical lattice: the smaller the modulation, the 
farther the dipole moves along the y-direction before annihilation, and the
longer it takes for the dipole to disappear.
This is shown in Fig.\ref{fig_core_1}, where it appears that the vortex hopping occur mostly
across bridge sites. The two vortices are initially placed at $x=\pm 2d=\pm 6.35\,\mu m$
(left panel of Fig.\ref{fig_core_1}).
In the right panel of Fig.\ref{fig_core_1} the annihilation paths are shown instead
when the two vortices are initially placed at a larger mutual distance, 
$x=\pm 3d=\pm 9.53\,\mu m$.
We recall that in absence of OL the vortex dipole
would move rigidly with constant velocity in the $y$ direction.

It is of interest a comparison with the dynamics of a vortex
dipole in the supersolid state of dipolar bosons \cite{Anc21}. There
is some similarity with what we find here in that the vortices
of the dipole move by approaching each other by jumps between
equilibrium sites with final annihilation but the studied case
for dipolar atoms did not show any translation of the dipole in the 
direction perpendicular to the line joining the vortices.


The energy released immediately after the annihilation
goes into excitations of the system. In order to gain some 
insights on the character of the excitations
we computed the spectral density of the
kinetic energy of the superfluid velocity field,
decomposing
it into compressible and incompressible parts \cite{tsubota,spectra,bradley}.
Briefly, one splits the density-weighted velocity field
$\vec u (\vec r,t)\equiv \sqrt{\rho (\vec r ,t)}\vec v (\vec r ,t)$
into a compressible (C) and an incompressible (I) part, 
$\vec u (\vec r,t)=\vec u ^I(\vec r,t)+\vec u ^C(\vec r,t)$,
such that $\vec \nabla \cdot \vec u ^I(\vec r,t)=0$ and
$\vec \nabla \times u ^C(\vec r,t)=0$.
One can therefore decompose the kinetic energy $E$ into two parts,
$E=E^I+E^C$, where

\begin{equation}
E^{I,C}=\frac {m}{2}\int d\vec r |\vec u^{I,C}(\vec r ,t)|^2
\end{equation}

The compressible component is attributed to the kinetic
energy contained in the sound field, while the incompressible 
part gives the contribution from quantum vortices.

In $\vec k$-space, the total incompressible (compressible) kinetic energy $E^{I,C}$ is
given by

\begin{equation}
E^{I,C}=\frac {m}{2}\Sigma _{j=x,y}\int d^2\vec k |F_j(\vec k)|^2
\end{equation}
where
\begin{equation}
F_j(\vec k)=\frac {1}{2\pi} \int d^2 \vec r e^{-i\vec k \cdot \vec r} u_j^{I,C}(\vec r)
\end{equation}
(the time-dependence is implied).

The one-dimensional spectral density in $k$-space is given by integrating 
over the azimuthal angle

\begin{equation}
E^{I,C}(k)= \frac {mk}{2} \Sigma _{j=x,y}  \int _0^{2\pi } d\phi _k |F_j(\vec k)|^2 
\end{equation}

\begin{figure}
\centerline{
\includegraphics[width=0.8\linewidth,clip]{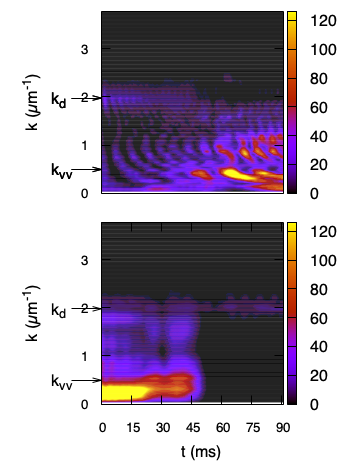}
}
\caption{Kinetic energy spectra for the case $\tilde{V}=4$, square OL. Top: compressible part;
bottom: incompressible part.  
}
\label{fig_spectra}
\end{figure}

We show in
Fig.(\ref{fig_spectra}) the spectral density of the kinetic energy, $E^{I,C}(k,t)$
for the case $\tilde{V}=4$, when the two vortices are initially placed 
at positions $x=\pm 2d$ (whose trajectories are displayed in the left panel 
of Fig.\ref{fig_core_1}.
The horizontal arrows show two relevant wave
vectors: $k_{vv} = 2\pi /l$, where $l$ is the initial vortex-vortex distance, and
$k_d = 2\pi/d$, which is the wave vector
corresponding to the periodic modulation with lattice constant $d$.
The lower panel clearly shows the sharp transition when the vortices 
disappear 
with a strong drop of the incompressible kinetic energy. The peak in
the incompressible part appearing
below the wave-vector $k_{vv}$
before annihilation is a general feature also
found in the calculations for a vortex dipole in Ref. \onlinecite{bradley}
(see in particular the Fig.3 of that reference).
After vortex annihilation the compressible part (upper panel)
starts showing features connected to density fluctuations (sound waves). One
can notice that also before vortex annihilation a faint time modulation is
present over an extended range of k vectors and its period is about 7 ms. We
have computed \cite{anc_next} the phonon frequencies of the Rb gas in presence of
the 2D OL but without vortices and find that this 7 ms period falls inside a gap
of the phonon spectrum. The vortex dipole represents a defect in the
modulated system and our interpretation of this oscillation is in term of a
localized vibration of the Rb gas around the moving vortex dipole.

We have also considered two other types of 2D lattices, with triangular symmetry
and with the honeycomb structure. Similarly with what we have done in the case of the
square lattice, we studied the vortex structures in such lattices and
the real-time dynamics of a vortex-antivortex pair, eventyally leading to 
annihilation as the vortices meet.
The results are reported in the Supplemental Material \cite{SM}.

\subsection{Visualizing vortices in optical lattices}

As previously discussed, 
the direct visualization of vortices in the system studied here 
might be difficult, especially for large values of the 
amplitude of the OL. This is a consequence of the fact that 
the vortex cores tend to localize in the low-density sites of the 
periodic lattice.
We suggest here that a simple way to visually detect the presence of vortices in
the modulated system can be achieved by sudden removal of the optical lattice potential.

One way to experimentally cause vortex nucleation is
by means of rotation of the optical lattice. 
We consider here the the case of the square lattice with the same radial confining
potential (circular "box") used to analyze the
dynamics of a single vortex in the 1-dimensional lattice described in Section III.C.

To enforce rotations (with some fixed angular velocity $\omega$,
around the z-axis) 
we work in the co-rotating frame, described by the Hamiltonian
\begin{equation}
\left\{H \,-\hbar \omega \hat{L}_z\right\} \,\Psi(\mathbf{r})  =  \,\mu \, \Psi(\mathbf{r})
\label{h_rot}
\end{equation}
where $\hat{\mathcal{L}}_z$ is the total angular momentum operator in the $z$-direction
and $H$ is the Hamiltonian of Eq.\eqref{eq:gpe}.

We first compute the stationary state in the presence of a rotation of the system
with constant $\omega $ by solving the Euler-Lagrange equations in imaginary time
associated with the previous Hamiltonian. If the imposed angular velocity
is large enough (i.e. larger than the critical frequency for a single 
vortex nucleation), a number of vortices will eventually populate the system, which may 
later be visualized. These vortices, as it happens in 
rotating finite superfluid samples, are initially nucleated on the 
boundaries of the system, and rapidly settle to an equilibrium position at 
a distance from the center which depends on the value of the rotational 
frequency: the larger the latter is, the closer to the center will be
the vortices in the stationary-state configuration.

\begin{figure}
\centerline{
\includegraphics[width=1.0\linewidth,clip]{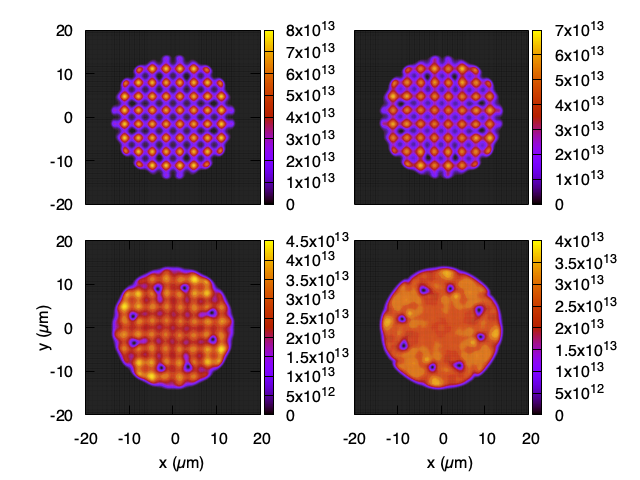}
}
\caption{From left to right, from top to bottom:
panel (1): density of the stationary state in the co-rotating frame,
for $\tilde{V}=7$;
panel (ii-iv): density patterns at $t=1.6$ ms, $t=4.1$ ms, and $t=6.4$ ms, 
during the time evolution of the state (1) when the amplitude $\tilde{V}$
is ramped linearly to zero in 5 ms.
The density is in units of $cm^{-3}$.
}
\label{rotation}
\end{figure}

As an example, we show in the first panel of Fig.(\ref{rotation}) 
the stationary state obtained with $\omega = 2\pi \times 8 $ Hz
in presence of a
square OL with value $\tilde{V}=7$.
At first sight, the resulting density profile resembles that 
of the ground-state in the absence of rotation.
However, a finite value for $\hat{\mathcal{L}}_z$ indicates the presence of vorticity
in the system.

A possible way to increase the vortex visibility
is to rapidly remove the
periodic potential (but keeping the radial confinement active)
so that the system may evolve towards a more homogeneous state
where the increased contrast between the empty core and the
background density may allow to reveal the vortex positions.

We therefore perform a real-time dynamics, starting from the
stationary state shown in the first panel of Fig.\ref{rotation},
with a linear ramp of the optical potential 
which brings its value from $\tilde{V}=7$ to zero in 5 ms.
The three panel (from left to right, from top to bottom) of Fig.\ref{rotation} show 
snapshots of the Rb density during the
real-time evolution, clearly showing a ring of six vortices as the density modulations
are suppressed.
Notice that the vortex core positions rotate with the imposed angular velocity $\omega $.

We remark that the time required to visually disclose the vortex cores 
is so short that the final core positions (last panel in the figure) 
coincide with their initial positions 
in the modulated phase shown in the first panel.
Such short times exclude the possibility that vortices are nucleated 
during the quench from the modulated to the homogeneous phase.

\bigskip

\section{Conclusions}
We have studied the superfluid phase of boson $^{87}$Rb atoms under the
influence of an optical potential which induces a spatial modulation of the
local density. The main interest is on the static and dynamical properties of
vortex excitations. We have studied the system at zero temperature with the
mean field Gross-Pitaevskii equation and our investigation is mainly on the
regime of strong modulation when the excursion between low and high
density becomes quite large. We study the system in a flat geometry so that
one can neglect the transverse direction and the system is close to the 2D
limit.
The dynamics of a vortex in an almost 2D superfluid is rather simple when no
other vortex is present within its healing length: the core moves with the
gradient of the phase due to other vortices. Consequences of this law are, for
instance, that a vortex dipole is a stable entity and it rigidly translates with
constant velocity and that one vortex in a circular trap performs a procession
around the trap center. Deviations from such behaviors in a weakly
inhomogeneous superfluid have been studied by a number of authors
as discussed in the Introduction. We have studied a rather different
regime, when the inhomogeneity is very large. In presence of an intense
optical potential we find a vortex behavior quite different from that of the
uniform superfluid. Depending on the symmetry of the optical lattice two
vortices of opposite chirality, a vortex dipole, can move by jumps approaching
each other until annihilation with a lifetime of the pair depending on the
intensity of the optical lattice or the vortices of the dipole do translate but with
a velocity even of opposite direction of that present in the uniform case and, in
addition, this translation takes place together with an oscillatory motion. Or a
single vortex in a trap does not perform a processional motion but it move
toward the periphery of the trap with a complex motion consisting of
translation and of an oscillation. These periodic motions are single-side in the
sense that they never cross the equilibrium position from the starting place, a
behavior quite different from that of a massive particle around an equilibrium
position. Such features of the vortex motion derive from two facts, on one
hand the vortex energy depends on the local density at the position of its core
and the energy is lowest where the density has a minimum, I.e. at the positions
of the maxima of the optical lattice. 

A consequence of this is that a vortex is
not free to move but it is pinned to specific sites or lines depending on the
symmetry of the optical lattice. On the other hand the stream lines have large
deviations from the simple circular shape of the uniform superfluid. The
superfluid fraction is reduced from unity, even if our system is at zero
temperature, and this reduction can be quite large for large amplitudes of the
optical potential. In the case of 1D optical potential the superfluid has a stripe
structure and the superfluidity is very anisotropic: the superfluid fraction is
unity along the stripes and it is reduced in the direction perpendicular to the
stripes. At the same time a vortex has a unit circulation but a reduced angular
momentum compared to that of a uniform superfluid.

It is possible now to generate experimentally vortex dipoles in a superfluid of
cold atoms \cite{kwon} so it should be possible to verify our predictions for the
dynamics of vortices in a modulated superfluid. We have studied the case of
$^{87}$Rb atoms but our results should be valid for other bosons with positive
scattering length.
The jumping behavior and annihilation of a vortex dipole seen in the case of a
2D optical lattice have some similarity with that of a vortex dipole in a
supersolid \cite{Anc21}.

Our study is based on a mean field theory and we should pose the question of the
accuracy of such theory because it is known that a Mott transition to a
localized state sets in when the optical lattice is strong enough \cite{greiner}. In terms
of recoil energy the amplitude of the optical potential in our study is much
smaller of the amplitudes for which experimentally the localization has been
found to set in. In addition we have indication of the internal consistency of
the used theory because we have performed computations \cite{anc_next} of the
excitation spectrum of our system with the Bogoliubov-de Gennes equation
and no sign of instability has been found. This gives confidence on our
theoretical results.

We have been able to explain at a qualitative level some aspects of our results
like the approach of the two vortices of a dipole or the one sided oscillations in
terms of the expression of the velocity of the vortex core and its relation to the
gradient of the local density of the superfluid. 

One would like to see a treatment of the systems of our study on the 
basis of an approximate analytic treatment like a suitable extension 
of the point-vortex approximation as explored in Ref. \onlinecite{simula}.
Different extensions of the present study come to mind. One is the study of
vortices in the supersolid phase of soft core bosons or in supersolid with a
stripe structure. The flat geometry used in our computations does not allow
flexural motion of the core of the vortex. Of interest will be a similar study
when the system is extended in the third direction and key questions are the
fate of the Kelvin waves that characterize the motion of a vortex core in 3D \cite{donnelly}
or if additional excitations are present similar to kinks of a dislocation in a
solid.
Theory predicts that a supersolid phase with a stripe structure can be present
with dipolar bosons \cite{ripley}. Our results for the vortex
behavior in a 1D modulated system should be relevant also for such supersolid system.

\begin{acknowledgments}
We thank Giacomo Roati, Alessia Burchianti and Andrea Richaud for useful discussions.
\end{acknowledgments}

\break
\newpage

{\bf Supplemental Material}
\bigskip

\section{Triangular and Honeycomb optical lattice.}

We consider here two other lattice types, with triangular symmetry.
The periodic potential  
mimics what is actually employed in experiment where a
potential with triangular symmetry is created by three laser beams that
intersect in the x–y plane mutually enclosing angles of 120$^\circ $ with 
lattice vectors $\vec k_1$, $\vec k_2$ and $\vec k_3$
(see Ref.\onlinecite{becker}). 

We use here the following form:
\begin{equation}
V_{OL}^{tri}=\tilde{V}\{ \frac {4}{3}-\frac {4}{9}[ cos(\vec b_1\cdot \vec R) + cos(\vec b_2\cdot \vec R)
+cos((\vec b_1-\vec b_2)\cdot \vec R) ] \}
\end{equation}
where $\vec R=(x,y)$, $\vec b_1=\vec k_2-\vec k_1$, $\vec b_2=\vec k_2-\vec k_3$,
and 
$\vec k_1=\kappa(\sqrt{3}/2,-1/2)$, $\vec k_2=\kappa (0,1)$, $\vec k_3=-\kappa(\sqrt{3}/2,1/2)$.
$k$ is the wave vector of the laser employed for the lattice beams. Here
$\kappa = 4\pi/(3d)$, $d$ being the lattice constant (the same used in the square
optical lattice).
The numerical coefficients in the previous equations 
are such to give the 
same min-max excursion, $(0,2V_0)$, of the square optical potential used in the previous sections. 
The number of Rb atoms $N$ is also chosen such 
to have the same mean density as in the square lattice case.
The density of the Rb atoms
is a periodic function of position with triangular symmetry.

A different symmetry of the density of the atoms can be obtained 
in the presence of the potential 

\begin{equation}
V_{OL}^{hon}(x,y)=-V_{OL}^{tri}(x,y)+2V_0
\end{equation}
so that the role of maxima and minima of $V_{OL}^{tri}(x,y)$ are exchanged.
In this case the maxima in the Rb density will be at the
points of an honeycomb lattice.

We have studied the vortex properties in these two cases.
As in the case of the square lattice we find that the
equilibrium positions of a vortex are at the sites of minimum
density.

\begin{figure}
\centerline{
\includegraphics[width=0.8\linewidth,clip]{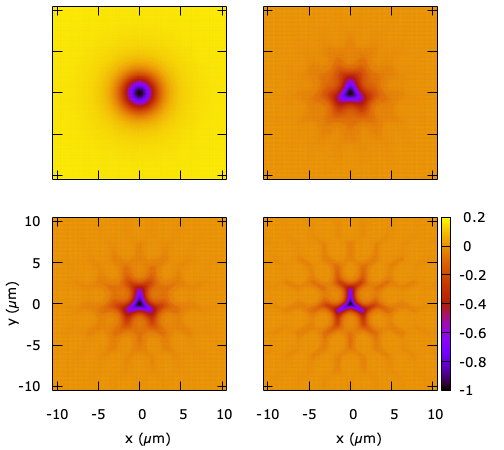}
}
\caption{Relative density difference $(\rho _{vort}-\rho _{novort})/\rho _{novort}$
showing the 
minimum energy vortex structure (hollow site) in the triangular lattice
for 
for $\tilde{V}=0,2,4,7$.
}
\label{fig4_tria}
\end{figure}

\begin{figure}
\centerline{
\includegraphics[width=0.8\linewidth,clip]{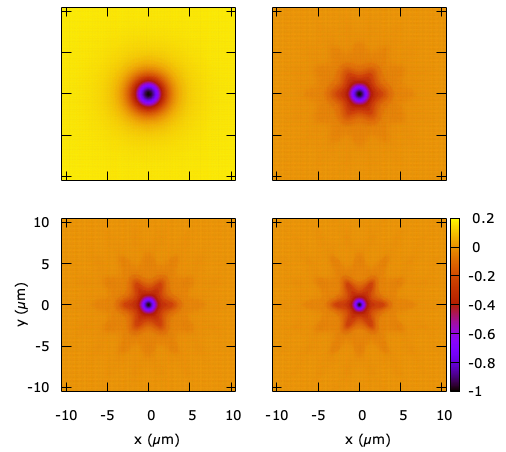}
}
\caption{Relative density difference $(\rho _{vort}-\rho _{novort})/\rho _{novort}$ 
showing the 
minimum energy vortex structure (hollow site) in the honeycomb lattice
for $\tilde{V}=0,2,4,7$.
}
\label{fig4_hexa}
\end{figure}

We show in Figs.(\ref{fig4_tria}) and (\ref{fig4_hexa}) the equilibrium vortex structure 
for different values of the potential depth $\tilde{V}$, including the value $\tilde{V}=0$,
in
the triangular and honeycomb lattice, respectively.
In order to facilitate the visualization 
we show the density differences
with respect to the configuration without vortex.

\begin{figure}
\centerline{
\includegraphics[width=0.9\linewidth,clip]{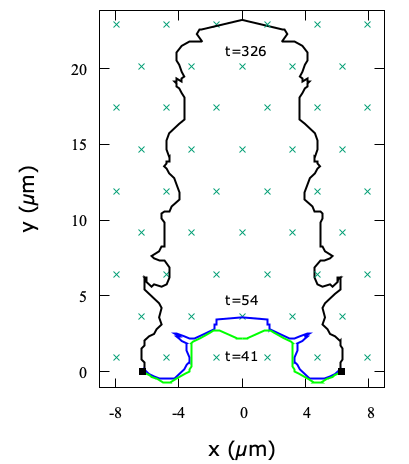}
}
\caption{Trajectories in the x-y plane ending in 
the vortex dipole annihilation, for the triangular lattice. 
The two vortices are initially placed at $x=\pm 2d=\pm 6.35\,\mu m$.
From top to bottom:
$\tilde{V}=2,4,7$. The label on each trajectory report the observed annihilation
time in milliseconds. The crosses show the positions of the T sites where the Rb density is maximum.
}
\label{fig_core_tria}
\end{figure}

\begin{figure}
\centerline{
\includegraphics[width=0.9\linewidth,clip]{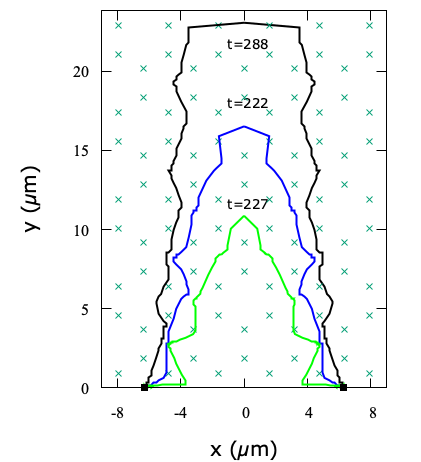}
}
\caption{Trajectories in the x-y plane ending in 
the vortex dipole annihilation, for the honeycomb lattice. 
The two vortices are initially placed at $x=\pm 2d=\pm 6.35\,\mu m$.
From top to bottom:
$\tilde{V}=2,4,7$. The label on each trajectory report the observed annihilation
time in milliseconds. The crosses show the positions of the T sites where the Rb density is maximum.
}
\label{fig_core_hexa}
\end{figure}

Finally, we show in Fig.(\ref{fig_core_tria}) and (\ref{fig_core_hexa})
the trajectories of a vortex-antivortex pair, obtained by solving the 
time-dependent GP equation, as they propagate through the 
lattice before undergoing annihilation after some time.
Again, the propagation occurs in the form of jumps across the 
lower energy barrier sites, and the lower the modulation, the longer
it takes for the annihilation to occur and the larger is the
displacement of the dipole before annihilation.
From these results and those for
the square lattice we conclude that the basic features of the motion of a vortex
dipole do not depend on the lattice symmetry and that the lifetime of the vortex
dipole can be controlled by the intensity of the OL.

\bigskip
\newpage




\begin{thebibliography}{99}

\bibitem{donnelly} 
R.J. Donnelly, {\it Quantized vortices in helium II},  Cambridge University Press (1991).

\bibitem{fetter} 
A.L. Fetter, Rev. Mod. Phys. {\bf 81}, 647 (2009).

\bibitem{olattice} S. Burger, F.S. Cataliotti, C. Fort, P. Maddaloni, F. Minardi and M. Inguscio,
Europhys. Lett. {\bf 57}, 1 (2002);
F.S. Cataliotti, S. Burger, C. Fort, P. Maddaloni, F. Minardi, A. Trombettoni, A.
Smerzi and M. Inguscio, Science {\bf 293}, 843 (2001).

\bibitem{tao} J. Tao, M. Zhao and I.B. Spielman,
Phys. Rev. Lett. {\bf 131}, 163401 (2023).

\bibitem{leg70} 
A.J. Leggett, Phys. Rev. Lett. {\bf 25}, 1543 (1970);
A.J Leggett, J. Stat. Phys. {\bf 93}, 927 (1998).

\bibitem{perez} D. Perez-Cruz, G.E. Astrakharchik and P. Massignan,
arXiv:2403.08416v1 (March 2024).

\bibitem{moroni} S. Moroni, F. Ancilotto, P. L. Silvestrelli and L. Reatto,
Phys. Rev. B {\bf 103}, 174514 (2021).

\bibitem{lchomaz} L. Chomaz, I. Ferrier-Barbut, F. Ferlaino, B. Laburthe-Tolra, B.L. Lev and T. Pfau,
Reports on Progress in Physics {\bf 86}, 026401 (2022).

\bibitem{isoshima} T. Isoshima, M. Nakahara, T. Ohmi, and K. Machida, Phys. Rev.
A {\bf 61}, 063610 (2000);
K. C. Wright, L. S. Leslie, A. Hansen, and N. P. Bigelow, Phys.
Rev. Lett. {\bf 102}, 030405 (2009).

\bibitem{neely} T.W. Neely, E.C. Samson, A.S. Bradley, M.J. Davis and B.P. Anderson,
Phys. Rev. Lett. {\bf 104}, 160401 (2010).

\bibitem{jlin} Y.-J. Lin, R.L. Compton, K. Jimenez-Garcia, J.V. Porto and
I.B. Spielman, Nature Letters {\bf 462}, 628 (2009).

\bibitem{kwon} W.J. Kwon, G. Del Pace, K. Xhani, L. Galantucci, A.M. Falconi, M. Inguscio, F. Scazza and G. Roati,
Nature {\bf 600}, 64–69 (2021).

\bibitem{kevrek} P.G. Kevrekidis et al., J. Phys. B: At. Mol. Opt. Phys. {\bf 36}, 3467 (2003).

\bibitem{goldbaum} D.S. Goldbaum and E.J. Mueller,
Phys. Rev. A {\bf 77}, 033629 (2008).

\bibitem{hpu} H. Pu, L.O. Baksmaty, S. Yi and N.P. Bigelow,
Phys. Rev. Lett. {\bf 94} 190401 (2005);
K. Kasamatsu and M. Tsubota, Phys. Rev. Lett. {\bf 97} 240404 (2006).

\bibitem{malomed} B.B. Baizakov, B. A. Malomed and M. Salerno, Europhys. Lett. {\bf 63}, 642-648 (2003);
B.B. Baizakov, B. A. Malomed and M. Salerno, Phys. Rev. E {\bf 74}, 066615 (2006).

\bibitem{kivshar} E.A. Ostrovskaya and Y.S. Kivshar, Phys. Rev. Lett. {\bf 93}, 160405 (2004);
T.J. Alexander, E.A. Ostrovskaya, A.A. Sukhorukov and Y.S. Kivshar,
Phys. Rev. A {\bf 72}, 043603 (2005).

\bibitem{epoli} E. Poli, T. Bland, S.J.M. White, M.J. Mark, F. Ferlaino, S. Trabucco and M. Mannarelli,
Phys. Rev. Lett. {\bf 131}, 223401 (2023).

\bibitem{gallemi} A. Gallemi, S.M. Roccuzzo, S. Stringari and A. Recati,
Phys. Rev. A {\bf 102}, 023322 (2020).

\bibitem{gallemi1} A. Gallemi and L. Santos,
Phys. Rev. A {\bf 106}, 063301 (2022).

\bibitem{casotti} E. Casotti, E. Poli, L. Klaus et al.,
arXiv:2403.18510v1 (2024).

\bibitem{Anc21}
F. Ancilotto, M. Barranco, M. Pi, and L. Reatto,
Phys. Rev. A {\bf 103}, 033314 (2021).

\bibitem{duine} J.W. Reijnders and R.A. Duine,
Phys. Rev. Lett. {\bf 93}, 060401 (2004);
S. Tung, V. Schweikhard and E. A. Cornell,
Phys. Rev. Lett. {\bf 97}, 240402 (2006);
H. Pu, L. O. Baksmaty, S. Yi, and N. P. Bigelow, Phys. Rev.
Lett. {\bf 94}, 190401 (2005).

\bibitem{marte} 
A. Marte, T. Volz, J. Schuster, S. Durr, G. Rempe, E.~G.~M. van Kempen, and B.~J. Verhaar, Phys. Rev. Lett. {\bf 89}, 283202 (2002).

\bibitem{Ral60} 
A. Ralston and H.~S. Wilf, {\it Mathematical methods for digital computers} (John Wiley and Sons, New York, 1960).

\bibitem{sep_joss_rica}
Y. Pomeau and S. Rica, Phys. Rev. Lett. {\bf 72}, 2426 (1994);
N. Sepulveda, C. Josserand, and S. Rica, Eur. Phys. J. B {\bf 78}, 439 (2010).

\bibitem{Pi07}
M. Pi, R. Mayol, A. Hernando, M. Barranco, and F. Ancilotto, J. Chem. Phys. {\bf 126}, 244502 (2007).

\bibitem{sadd} 
M. Sadd, G. V. Chester, and L. Reatto, Phys. Rev. Lett. 79, 2490 (1997).

\bibitem{Anc18}
F. Ancilotto, M. Barranco and M. Pi, Phys. Rev. B {\bf 97}, 184515  (2018).

\bibitem{simula} A.J. Groszek, D.M. Paganin, K. Helmerson and T.P. Simula,
Phys. Rev. A {\bf 97}, 023617 (2018).

\bibitem{roberts} 
C.A. Jones and P.H. Roberts, J. Phys. A {\bf 15}, 2599 (1982); A. Griffin, 
V. Shukla, M.-E. Brachet, and S. Nazarenko, Phys. Rev. A {\bf 101}, 053601 (2020).

\bibitem{anc_next} F. Ancilotto, unpublished.

\bibitem{tsubota} M. Tsubota, K. Fujimoto, S. Yui, J. Low. Temp.
Phys. {\bf 188}, 119 (2017).

\bibitem{spectra} C. Nore, M. Abid, and M.E. Brachet, Phys. Fluids {\bf 9}, 2644 (1997).

\bibitem{bradley} A.S. Bradley and B.P. Anderson, Phys. Rev. X {\bf 2}, 041001 (2012).

\bibitem{SM}
See Supplemental Material at  [{\it URL will be inserted by publisher}].

\bibitem{greiner} M. Greiner, M.O. Mandel, T. Esslinger, T. Hansch and I. Bloch, 
Nature {\bf 415}, 39 (2002).

\bibitem{richaud} A. Richaud, V. Penna and A. L. Fetter,
Phys. Rev. A {\bf 103}, 023311 (2021).

\bibitem{richaud1} A. Bellettini, A. Richaud and V. Penna, Phys. Rev. A {\bf 109}, 053301 (2024).

\bibitem{ripley} B.T.E. Ripley, D. Baillie and P.B. Blakie, Phys. Rev. A {\bf 108}, 053321 (2023).

\bibitem{becker} C. Becker, P. Soltan-Panahi, J. Kronjager, S. Dorscher,
K. Bongs and K. Sengstock,
New Journal of Physics {\bf 12}, 065025 (2010).

\end{thebibliography}
\end{document}